\documentclass[aps,prl,twocolumn,amsmath, amssymb,nofootinbib,superscriptaddress, floatfix]{revtex4-1}
\usepackage{color}
\usepackage{graphicx}
\def\beq{\begin{equation}}
\def\eeq{\end{equation}}
\def\bea{\begin{eqnarray}}
\def\eea{\end{eqnarray}}
\def\avg#1{\left\langle#1\right\rangle}

\def\be{\begin{equation}}       \def\ee{\end{equation}}
\def\bea{\begin{eqnarray}}      \def\eea{\end{eqnarray}}
\def\ba{\begin{array} }
\def\ea{\end{array} }
\def\bnum{\begin{enumerate} }
\def\enum{\end{enumerate}}

\def\nn{\nonumber}
\def\pa{\partial}
\def\=>{\Rightarrow}
\def\>{\rightarrow}
\def\A{\uparrow}
\def\V{\downarrow}

\def\eye2{Fathbb{I}}

\begin{document}

\title{Interaction effect on topological classification of superconductors
in two dimensions}

\author{Hong Yao}
\affiliation{
Institute for Advanced Study, Tsinghua University, Beijing, 100084, China}
\affiliation{
Department of Physics, Stanford University, Stanford, California 94305, USA}

\author{Shinsei Ryu}
\affiliation{
Department of Physics, University of Illinois at Urbana-Champaign,
1110 West Green St, Urbana, IL 61801, USA}

\date{\today}

\begin{abstract}
We introduce a new class of superconductors (SCs) in two spatial dimensions with time reversal symmetry and reflection (i.e., mirror) symmetry.
In the absence of interactions, topological classes of these SCs
are distinguished by an integer-valued ($\mathbb{Z}$) topological invariant.
When interactions are included, we show that the topological classification
is modified to $\mathbb{Z}_8$. This clearly demonstrates that interactions can have qualitative effect on topological classifications of gapped states of matter in more than one dimension.
\end{abstract}

\maketitle

{\bf Introduction}:
Topological phases are fully quantum mechanical
states of matter which are not characterized by classical symmetry
breaking\cite{wen89}. While gapped in the bulk, quite often, they are accompanied by {\it gapless}
excitations at their boundary, signaling highly entangled nature
of their ground state.
Since the discovery of the integer quantum Hall effect (IQHE)
\cite{review_QHE},
the list of topological phases in nature
has been expanded,
in particular, by
the recent discovery of topological insulators
in two and three dimensions (2D and 3D)
in systems with strong spin-orbit coupling
\cite{review_TIa, review_TIb,KaneMele,Roy06, Bernevig05,Moore06,Roy3d,Fu06_3Da,Fu06_3Db,qilong},
and the identification of $^{3}$He B as a topological SC
(superfluid)
\cite{3HeB}.
Unlike the IQHE,
the topological character of these topological insulators
and SCs
(i.e., the stable gapless edge or surface modes)
is protected by time-reversal symmetry (TRS).
The presence or absence of a topological distinction among gapped phases
for a given set of symmetries and for given spatial dimensions
can be studied systematically, and
is summarized in the ``periodic table'' of topological insulators and SCs for non-interacting \cite{footnote1} fermions
\cite{Schnyder2008,Kitaev2009,SRFLnewJphys}.

Since interactions are ubiquitous in real materials, a natural question
is whether and how interactions
could modify topological classifications
obtained for non-interacting systems\cite{wu}.
In other words, are there cases where interactions qualitatively
affect topological classifications?
Even though this question has not been fully or thoroughly answered,
an exciting progress has been made
in Ref.\ \cite{Fidkowski-Kitaev},
in which
Fidkowski and Kitaev
explicitly showed
that the putative $\mathbb{Z}$ classification
of one-dimensional (1D) ``non-interacting'' SCs
with unusual time reversal symmetry
(``BDI class'')
is modified to $\mathbb{Z}_8$ when interactions
are included.
This $\mathbb{Z}_8$ classification of the 1D BDI SCs
with interactions is further illustrated from the study
of entanglement spectrum
\cite{Fidkowski2,Turner}.
In more than one dimension, stability of topological classifications
in a number of classes of non-interacting topological insulators
and SCs is argued from the perspective of
low-energy topological response theory
\cite{qilong,Ryu-Moore-Ludwig,Wang-Qi-Zhang}. For bosonic (or spin) systems, a systematic approach of constructing symmetry protected topological phases in general spatial dimensions was recently proposed
\cite{Chen}. However, how interactions in fermionic systems in 2D and 3D can dramatically modify their topological classification obtained from non-interacting fermions remains largely unexplored \cite{gu}.

In this paper, we partly fill this gap by considering 2D SCs with TRS ($\mathcal{T}^2=-1$) and reflection symmetry (RS) ($\mathcal{R}^2=-1$).
Note that $\mathcal{T}$ is anti-unitary while $\mathcal{R}$ is unitary.
We call SCs with these symmetry properties
``DIII+R'' SCs. First, we show that the topological classes of these DIII+R SCs are classified by an integer-valued ($\mathbb{Z}$) topological invariant at the quadratic level. Such topologically non-trivial SCs are characterized by helical Majorana modes on the system's edge. The number of helical Majorana edge states is protected by the symmetries at the non-interacting level.

We then ask whether those helical gapless Majorana edge modes are stable against interactions while preserving those relevant symmetries in question.
As we show later in detail, the helical Majorana edge states
in the case of $\mathbb{Z}=8$ topological invariant
are unstable against (even weak) interactions
-- the gapless helical Majorana edge states become gapped
while no symmetry is broken in the bulk or the edges of the system.
It is worth to stress that all the relevant symmetries are
fully preserved while the gap opens in the edge.
In other words, the edge of the putative $\mathbb{Z}=8$ system
is qualitatively the same as $\mathbb{Z}=0$ systems
when symmetry-preserving interactions are included. Because the system's edge we consider respects the same set of symmetries as its bulk, robust edge states could fully encode topological properties of the bulk due to the bulk-edge correspondence. Consequently, the putative $\mathbb{Z}$ topological classification of non-interacting ``DIII+R'' superconductors in 2D is reduced $\mathbb{Z}_8$ when interactions are considered.
A similar model, but with different set of symmetries,
was studied independently in Refs. \cite{XLQ2012,RyuZhang2012}.

{\bf DIII+R SCs}:
Two-dimensional SCs with TRS $\mathcal{T}^2=-1$
(symmetry class DIII)
have a
$\mathbb{Z}_2$ topological invariant as shown in the ``periodic table''.
To see this explicitly, it is helpful to
study their
edge theory.
For a $\mathbb{Z}_2$ non-trivial DIII SCs in 2D,
its
generic edge theory is the 1D helical Majorana fermions:
\bea
H_{\rm{free}}=\int dx \left[ \psi_{\A}i\pa_x \psi_{\A} - \psi_{\V} i\pa_x \psi_{\V}\right],
\label{edge Hamiltonian}
\eea
where $\psi_{\sigma}$ are left/right-moving edge Majorana fermion operators
and $\sigma=\A,\V$ are
spin index
with the following properties under time reversal transformation:
$\mathcal{T}^{-1} \psi_{\A} \mathcal{T} =\psi_{\V}$ and $\mathcal{T}^{-1}\psi_{\V}\mathcal{T}=-\psi_{\A}$,
which satisfies $\mathcal{T}^2=-1$.
It is clear that the helical edge states above respect the TRS.
Moreover,
this TRS
protects this putative gapless helical edge state since there is only one mass term $im\psi_{\A}\psi_{\V}$
which breaks
the TRS
in the edge. However, when there are two such helical gapless modes labeled by $a=1,2$, the putative gapless modes can be fully gapped by time reversal invariant terms
$im(\psi^1_{\A}\psi^2_{\V} +\psi^1_{\V}\psi^2_{\A})$.
Consequently, the topological classification is $\mathbb{Z}_2$.

To possibly have a
$\mathbb{Z}$ topological classification of DIII SCs in 2D,
further discrete symmetry is needed. Since the helical edge states
preserve not only TRS but also the RS
$\mathcal{R}$ defined as
 $\mathcal{R}^{-1}\psi_{\A}(x)\mathcal{R} =\psi_{\V}(-x)$
 and
 $\mathcal{R}^{-1}\psi_{\V}(x) \mathcal{R}=-\psi_{\A}(-x)$
\cite{footnote4},
we consider 2D DIII SCs with RS, dubbed as the DIII+R class.
Note that $\mathcal{R}^2=-1$ as required for spin half-integer fermions.
Now, we show that the DIII+R SCs in 2D are
distinguished by an integer-valued ($\mathbb{Z}$) topological invariant.
To see this, we consider $N$ copies (or flavors) of the helical gapless Majorana edge states given by
$\sum_{a=1}^N\int dx\, v_a
\big[i\psi^a_{\A}\pa_x\psi^a_{\A} -i\psi^a_{\V}\pa_x\psi^a_{\V} \big],
\label{N edge Hamiltonian}$
where $a=1,\ldots,N$ are the ``flavor'' indices (which is preserved by both the time-reversal or reflection transformations) and $v_a$ are the Fermi velocities.
We then write down the most general mass term
\bea
i \psi_{\A}^a M_{ab} \psi_{\V}^b,
\quad
M^*_{ab} = M_{ab},
\eea
and check if it can preserve both TRS and RS.
(Note that terms proportional to $i\psi_{\sigma}^a \psi_{\sigma}^b$ for $\sigma=\A$ or $\V$ are irrelevant and only renormalize the Fermi velocities of edge Fermions.)
Time-reversal symmetry requires $M_{ab}=M_{ba}$ while
RS requires $M_{ab}=-M_{ba}$.
Consequently,
$M=0$ identically.
In other words, it is impossible to gap the $N$ pairs of helical gapless Majorana
edge states by considering the non-interacting fermion-bilinear terms for arbitrary $N$.
This class of SCs is then characterized by an integer-valued ($\mathbb{Z}$) topological invariant (see below).

{\bf Microscopic models}:
Armed with insights from the edge theories,
we now introduce an explicit mean-field SC Hamiltonian which respects both
TRS and RS and which has non-trivial gapless helical Majorana edge states:
\bea
H_{\rm{latt.}}&=&
\sum_{\avg{ij}\sigma}
\big[
-tc^\dag_{i\sigma} c_{j\sigma} +H.c.
\big]
-\mu
\sum_{i \sigma}
c^\dag_{i\sigma} c_{i\sigma}
\nn\\
&+&
\sum_i
\big[
\Delta(c^\dag_{i\A}c^\dag_{i+\hat x\A}+c^\dag_{i\V}c^\dag_{i+\hat x\V})+H.c.
\big]
\nn\\
&+&
\sum_i
\big[
{i} \Delta( c^\dag_{i\A}c^\dag_{i+\hat y\A}-c^\dag_{i\V}c^\dag_{i+\hat y\V})+H.c.
\big],
\eea
where $c^\dag_{i\sigma}$ are the fermion creation operators on site $i$ and $\sigma=\A,\V$ are spin indices. It is clear that the model above is
invariant under RS $\mathcal{R}$:
$c_{x,y,\sigma}\to (i\sigma^x)_{\sigma\sigma'} c_{-x,y,\sigma'}$
and TRS
$\mathcal{T}$:
$c_{i\sigma}\to  (i\sigma^y)_{\sigma\sigma'} c_{i\sigma'}$
\cite{footnote4}.
This model describes a SC with spin up $p_x+ip_y$ pairing and spin down $-p_x+ip_y$ pairing, and can describe a thin film of $^{3}$He-B.
On a cylinder with edges parallel to the $x$-direction,
we (numerically) obtain the helical Majorana edge states,
for $t=\Delta=1$ and for $-4<\mu<0$ or $0<\mu<+4$, say.
These helical edge states are effectively described
by the Hamiltonian (\ref{edge Hamiltonian}).
The $N$-flavor generalization of this model is straightforward.

{\bf Bulk topological invariant}:
There is a bulk
topological
invariant which guarantees,
when non-interacting and when there is a translation symmetry,
the stability of the edge modes of DIII+R topological SCs
for arbitrary $N$.
The construction is similar in spirit to the mirror Chern-number in 3D topological insulators protected by
RS \cite{TeoFuKane2008}.
With the periodic boundary condition,
the quadratic bulk Hamiltonian can be Fourier transformed as
\begin{align}
&
H
=
\sum_{0\le k_x \le \pi}
\sum_{k_y}
\Psi^{\dag}_{k_x}(k_y)
\mathcal{H}_{k_x}(k_y)
\Psi^{\ }_{k_x}(k_y),
\nonumber \\
&\Psi^{\dag}_{k_x}(k_y)
: =
\big(
\begin{array}{cccc}
c^{\dag}_{\uparrow,k}, &
c^{\dag }_{\downarrow,k}, &
c^{\ }_{\uparrow,-k}, &
c^{\ }_{\downarrow,-k}
\end{array}
\big).
\end{align}
We then note
at the reflection symmetric points
$k_x=\tilde k_x$($=0$ and $\pi)$,
the Bloch Hamiltonian $\mathcal{H}_{\tilde k_x}(k_y)$ commutes
with
$\mathcal{R}$,
namely $[\mathcal{H}_{\tilde k_x}(k_y),J^x]=0$, where $J^x=\textrm{diag}(i\sigma^x,-i\sigma^x)$.
In other words, at $\tilde k_x$, the quadratic Hamiltonian conserves $S^x$. Combined with TRS,
the quadratic Hamiltonian at $k_x=\tilde k_x$ can be written as
\begin{eqnarray}
H(\tilde k_x)=
\sum_{k_y}
(c^\dag_{+,k},
c^{\ }_{-,-k})
\bar{\mathcal{H}}_{\tilde k_x}(k_y)
\left(
\begin{array}{c}
c^{\ }_{+,k} \\
c^\dag_{-,-k}
\end{array}
\right),\quad
\end{eqnarray}
where the subscripts $\pm$ are eigenvalues of $\sigma^x$.
Now, the Hamiltonian
$\bar{\mathcal{H}}_{\tilde k_x}(k_y)$
above is in the AIII class with chiral symmetry
\cite{Schnyder2008}.

Following Ref.\ \onlinecite{Schnyder2008},
gapped 1D quadratic Hamiltonians in symmetry class AIII
are distinguished by an integer topological invariant, the winding number $\nu$.
Thus, at each reflection symmetric momentum $k_x=\tilde k_x$ ($\tilde k_x=0$ and $\pi$),
we can introduce an integer topological invariant
(``the reflection winding number''), $\nu(\tilde k_x)$,
the winding number of
$\bar{\mathcal{H}}_{\tilde k_x}(k_y)$.
We thus have two integral topological invariants,
$\tilde{\nu}(k_x=0,\pi)$.
The non-zero value of the invariant, $\tilde{\nu}(\tilde k_x)\neq 0$,
guarantees the presence of $|\tilde{\nu}(k_x)|$ pairs
of
zero-energy Majorana states at $\tilde k_x$,
when an edge is introduced along $x$-direction.
In particular,
when the invariant is non-zero $\tilde{\nu}(\tilde k_x)\neq 0$
at one of the reflection symmetric momenta ($\tilde k_x=0$, say)
and it is zero at the other ($\tilde k_x=\pi$),
this means there must be $|\tilde{\nu}(k_x)|$ branches of
non-chiral edge modes.
In general,
the difference
of the reflection winding number
\begin{align}
\tilde{\nu}(k_x=0)
-
\tilde{\nu}(k_x=\pi)
\end{align}
tells us the number of non-chiral edge modes.

{\bf Interaction effect}:
At the non-interacting level (and without disorder),
the DIII+R SCs have
$\mathbb{Z}$ topological classification as shown from both bulk and edge theories. Now, we consider the effect of symmetry-preserving interactions in the mean-field BdG Hamiltonian and check if the putative topological classification of $\mathbb{Z}$ is modified or not.
Since gapless helical edge states are the hallmark of those topologically
nontrivial SCs, we believe that it would be sufficient to check
if the gapless helical edge states are stable against interactions
while requiring there is no symmetry breaking induced by interactions
\cite{footnote3}.

For the case of
topological invariant $N=8$ [more generally $N\equiv 0$ (mod 8)], we try to identify certain interactions that
can destabilize the gapless edge states while preserving the symmetries of the bulk and the edge. For simplicity, we consider the helical edge modes of free fermions with the same Fermi velocities $v_a=v$
\bea
H_{\rm{free}}=v\int dx \sum_{a=1}^8 \left[\psi^a_{\A}i\pa_x \psi^a_{\A} -\psi^a_{\V}i\pa_x\psi^a_{\V}\right],
\label{free}
\eea
which is invariant under the global SO(8) rotations among
the left-moving or right-moving Majorana fermions.
Now, we consider interactions allowed by TRS and RS.
One naturally starts with the following SO(8) symmetric interactions
$
H_{\rm{GN}}
=
H_{\rm{free}}
-g \int dx  \big(\sum_{a=1}^8 i\psi^a_{\A}\psi^a_{\V}\big)^2
$,
where $g$ is the coupling constant.
This theory is then the SO($8$) Gross-Neveu (GN) model in (1+1)D,
which is exactly solvable \cite{shankar}.
Especially, the interaction is marginally relevant for $g>0$:
for arbitrary small interaction strength $g$
the ground state is gapped by spontaneously breaking
the
time reversal (or $\mathbb{Z}_2$ chiral) symmetry
with the order parameter
$\langle i\psi^a_{\A}\psi^a_{\V} \rangle \sim e^{-\pi/(vg)}$.
There are twofold degenerate ground
states at the edge.
When $g<0$, the interaction is marginally irrelevant and the ground state
remains gapless. In other words, the SO(8) symmetric GN interactions
cannot result in a unique gapped ground state in the edge.
We need to look for
some other channel of interactions
to fulfill this.

We follow the construction
introduced in Ref. \cite{Fidkowski-Kitaev}.
The non-interacting edge is described
by the conformal field theory (CFT) of 8 free Majorana fermions,
which is equivalent to the SO(8)$_1$
Wess-Zumino-Witten
model. The $8$ fermion operators $\psi^a_\sigma$ ($\sigma=\A$ or $\V$)
form the vector representation of SO(8).
Moreover, the 16-dimensional spinor representation of SO(8)
is reducible to two 8-dimensional irreducible ones formed by
spinor operators $\eta^a_\sigma$ and $\chi^a_\sigma$, $a=1,\ldots,8$.
The explicit forms of $\eta^a_\sigma$ and $\chi^a_\sigma$ are given by
\bea
\exp
\left[
\frac{i}2 (\pm\phi^1_\sigma\pm \phi^2_\sigma \pm\phi^3_\sigma \pm\phi^4_\sigma)
\right]
\label{def}
\eea
where $\phi^a_\sigma$ are boson fields obtained from bosonizing
the system,
$\psi^{2a-1}_\sigma\pm i\psi^{2a}_\sigma=e^{\pm i\phi^a_\sigma}$. The number of minus sign in the exponent of
Eq.\ (\ref{def}) is even for $\eta^a$ but odd for $\chi^a$.
Accidently, $\psi$, $\eta$ and $\chi$ all form 8-dimension representations of SO(8); they can actually be transformed into one another by the so-called triality symmetry of
SO(8).
It turns out that the spinor fields are useful to construct the interactions we desire.

To fully gap the edge states without spontaneously breaking any symmetry, we consider the following interactions
\begin{align}
H_{\rm{int}}
&=
-\int dx\left[
A
\left(\sum\nolimits_{a=1}^7i\eta^a_{\A}\eta^a_{\V}
\right)^2
\right.
\nonumber \\
&
\qquad \qquad
\left.
+B
\left(
\sum\nolimits_{a=1}^7 i\eta^a_{\A}\eta^a_{\V}\right)
\big(i\eta^8_{\A}\eta^8_{\V}\big)\right],
\end{align}
which is SO(7)-invariant and leaves $\eta^8_\sigma$ fixed.
This SO(7)-symmetric interaction is also local in terms of the original fermions $\psi^a_\sigma$.
Indeed, a finite gap opens in the edge states while preserving the symmetries in question, as explicitly shown in Ref. \cite{Fidkowski-Kitaev} when $B<0$ and $2A>B$. To understand this,
let us
first look at the limit $A\gg|B|$ which is the SO(7) GN model plus free $\eta^8_\sigma$ fermions (when $B\to 0$). The chiral symmetry is broken by the SO(7) GN interactions with the order parameter
$M=\langle i\sum_{a=1}^7 i\eta^a_{\A}\eta^a_{\V} \rangle \neq 0$, which generates a mass term $iBM\eta^8_{\A}\eta^8_{\V}$ for $\eta^8$ fermions. Now, the $\eta^8$ fermions can be mapped to the transverse field Ising model with field strength $(h-h_c)\propto \pm BM$, where $h_c$ is the critical field strength in the transverse field Ising model which is equal to the Ising interaction strength. For $h>h_c$, its ground state is paramagnetic without any symmetry breaking; for $h<h_c$, the system spontaneously breaks the Ising symmetry resulting in twofold degenerate ground states. In other words, $B>0$ and $B<0$ lie in two different phases. Since $B=2A>0$ is the SO(8) GN model which spontaneously break the chiral symmetry having twofold degenerate ground states, it is then clear that $B<0$ phase has a unique gapped ground state without breaking any symmetry.

It is worthwhile to understand more
heuristically why $N=8$ is special.
The edge theory with $N=8$
is qualitatively equivalent to the two-leg ladder electron model
at half-filling. Since there are two electrons per unit cell for
the two-leg ladder at half-filling,
having a fully gapped ground state without breaking any symmetry
is possible and expected
\cite{Lin-Balents-Fisher}.
To illustrate this, let us consider
lattice Majorana fermions on 1d chains described by the Hamiltonian
\cite{Fidkowski-Kitaev},
$
H= u H_1 + w H_2
$
where $H_1$ is quadratic in
lattice real fermion operators:
$
H_1
=
-({i}/2)
\sum_{a=1}^{8}
\sum_j
s_a \lambda^a_{j} \lambda^a_{j+1},
$
where
$\{\lambda^a_i,\lambda^b_j\}=2\delta^{ab}_{ij}$ and
$s_a=\{1,1,-1,-1,1,1,-1,-1\}$.
It is clear that $H_1$ is invariant under either TRS or RS, when we assume
\begin{align}
\mathcal{R}:&~\lambda^a_j\to (-1)^a \lambda^{a+2}_{-j}, \lambda^{a+2}_j\to -(-1)^a\lambda^a_{-j},\\
\mathcal{T}:&~\lambda^a_j\to\lambda^{a+2}_j, ~\lambda^{a+2}_j\to-\lambda^a_j,~~\quad\quad a=1,2,5,6.
\end{align}
$H_2$ is an interaction, and given by
$
H_2 = \sum_j
W(\lambda^a_{j})
$
where $W(\lambda^{a}_j)$ is a four fermion interaction composed of
eight Majorana fermions:
\begin{align}
&W(\lambda^a)
=
+\lambda^1 \lambda^2 \lambda^3 \lambda^4
+\lambda^1 \lambda^2 \lambda^5 \lambda^6
+\lambda^1 \lambda^2 \lambda^7 \lambda^8
 \\
&\quad
+\lambda^3 \lambda^4 \lambda^5 \lambda^6
+\lambda^3 \lambda^4 \lambda^7 \lambda^8
+\lambda^5 \lambda^6 \lambda^7 \lambda^8
\nonumber \\
&\quad
-\lambda^2 \lambda^3 \lambda^6 \lambda^7
-\lambda^1 \lambda^4 \lambda^5 \lambda^8
+\lambda^1 \lambda^3 \lambda^5 \lambda^7
+\lambda^2 \lambda^4 \lambda^6 \lambda^8
\nonumber \\
&\quad
-\lambda^2 \lambda^3 \lambda^5 \lambda^8
-\lambda^1 \lambda^4 \lambda^6 \lambda^7
-\lambda^1 \lambda^3 \lambda^6 \lambda^8
-\lambda^2 \lambda^4 \lambda^5 \lambda^7.\nn
\end{align}
In the absence of the interaction term
$w=0$, the lattice model is gapless,
whose continuum limit is given by
the Hamiltonian (\ref{free}).
When we switch on $w\neq 0$,
the edge theory is gapped with unique ground state
without breaking symmetries.
This can be understood in the following way:
the interaction $W$ can be written in terms of complex fermions
\begin{align}
(\lambda^1 + {i} \lambda^2 )/2
=
c^{\ }_{1\uparrow},
\quad
(\lambda^3 - {i} \lambda^4)/2
=
c_{1\downarrow},
\nonumber \\
(-\lambda^5 + {i} \lambda^6)/2
=
c^{\ }_{2\uparrow},
\quad
(\lambda^7 + {i} \lambda^8)/2
=
c_{2\downarrow},
\end{align}
as follows:
\begin{align}
W(\lambda_j^a)
=
16
\boldsymbol{S}_{j,1} \cdot \boldsymbol{S}_{j,2}
+
2 (n_{j,1}-1)^2
+
2 (n_{j,2}-1)^2 -2,
\end{align}
where
$\boldsymbol{S}_{1,2}$
and
$n_{1,2}$
is the spin
and
the fermion number operator
for $c_{1,2 s}$ ($s=\uparrow/\downarrow$).
With this interaction,
the charge degrees of freedom will be frozen by
Mott physics.
The exchange interaction
$\boldsymbol{S}_1 \cdot \boldsymbol{S}_2$
(the ``rung-exchange'' interaction)
realizes
the rung-single phase
which has a unique ground state without breaking any symmetry.

{\bf Discussion}:
We have discussed
interaction effects on
topological SCs
in 2D
protected by TRS and RS.
This is a non-trivial example in 2D
where topological classification of ground states
are dramatically altered by the interaction effect.
In a separate paper,
we plan on a more systematic study
on topological insulators and SCs protected by
RS in addition to other possible discrete symmetries
\cite{YaoRyu2012b,Bernard2012}.

We close with discussion on disorder
effects:
In the ten-fold classification of topological insulators and SCs,
it has been
proved useful to consider the boundary (edge, surface, etc.)
Anderson localization problem:
For a topological bulk, one should find
a boundary mode which is completely immune to disorder.
In turn, once one finds such
``Anderson delocalization'' at the boundary,
it means there is a topologically non-trivial bulk.
Not only this bulk-boundary correspondence can be used to
find and classify bulk topological phases
{\it in the absence of disorder},
it immediately tells us such topological phases are stable
against disorder.
For topological phases protected by a set of spatial symmetries,
stability against disorder is, in general, not trivial, since
spatial inhomogeneity does not respect the spatial symmetries.
One can still consider, however, situations where
the spatial symmetries are preserved {\it on average}.
The effects of disorder in $N$-channel quantum wires
in symmetry class DIII,
which are from our point of view the edge theory of
the DIII+R topological SC with the topological integer ($=N$),
have been studies
\cite{Brouwer2005}.
It is known that there is an even-odd effect in $N$:
the mean conductance
decreases algebraically as $L^{-1/2}$
with the length of the wire $L$ for odd $N$,
whereas it decays exponentially with $L$ for even $N$.
Correspondingly, for symmetry class DIII with $N$ odd
the density of states shows
the Dyson singularity.
This implies that
disorder simply reduces
the $\mathbb{Z}$ topological classification of DIII+R SCs
to
the $\mathbb{Z}_2$ classification,
which is the same as
the topological classification of 2D DIII SCs
in the periodic table.
In a separate paper
\cite{YaoRyu2012b}
we will report the other cases
where the $\mathbb{Z}$ topological classification
is reduces to the $\mathbb{Z}_2$ classification,
which is not related to the existing topological class
in the periodic table.

Note added: Recently, some other papers, a couple of works
that deal with similar topic have appeared\cite{sato,kane,furusaki}.

{\bf Acknowledgements:}
We would like to thank
Eun-Ah Kim,
Steve Kivelson, Dung-Hai Lee, and Shou-Cheng Zhang for helpful discussion,
and in particular Xiao-Liang Qi for sharing his results with us prior to its arXiv submission. This work is supported in part by Tsinghua Startup Fund and by US NSF Grant DMR-0904264.


\begin{thebibliography}{99}

\bibitem{wen89} X.-G. Wen, Phys. Rev. B {\bf 40}, 7387 (1989).

\bibitem{review_QHE}
{\it The Quantum Hall Effect},
edited by R. E. Prange and S. M. Girvin (Springer, New York, 1987).

\bibitem{review_TIa}
M. Z. Hasan and C. L. Kane, Rev. Mod. Phys. {\bf 82}, 3045 (2010).

\bibitem{review_TIb}
X.-L. Qi and S.-C. Zhang, Rev. Mod. Phys. \textbf{83}, 1057 (2011).

\bibitem{KaneMele}
C. L. Kane and E. J. Mele,
Phys. Rev. Lett. \textbf{95}, 146802 (2005);
C. L. Kane and E. J. Mele,
Phys. Rev. Lett. \textbf{95}, 226801 (2005).

\bibitem{Bernevig05}
B. A. Bernevig and S.-C. Zhang,
Phys. Rev. Lett. \textbf{96}, 106802 (2006).

\bibitem{Roy06}
R. Roy,
Phys. Rev. B \textbf{79}, 195321 (2009).

\bibitem{Moore06}
J. E. Moore and L. Balents,
Phys. Rev. B \textbf{75}, 121306(R) (2007).

\bibitem{Fu06_3Da}
L. Fu, C. L. Kane, and E. J. Mele,
Phys. Rev. Lett. \textbf{98}, 106803 (2007).

\bibitem{Roy3d}
R. Roy,
Phys. Rev. B \textbf{79}, 195322 (2009).

\bibitem{Fu06_3Db}
L. Fu and C. L. Kane,
Phys. Rev. B \textbf{76}, 045302 (2007).

\bibitem{qilong}
X.-L. Qi and T. Hughes and S.-C. Zhang,
Phys. Rev. B \textbf{78}, 195424 (2008).


\bibitem{3HeB}
See,
Y. Wada \textit{et al.},
Phys.\ Rev.\ B \textbf{78}, 214516 (2008),
S. Murakawa \textit{et al.},
Phys.\ Rev.\ Lett.\ \textbf{103}, 155301 (2009),
S.\ Murakawa {\it et al.},
J.\ Phys.\ Soc.\ Jpn.\ \textbf{80}, 013602 (2011),
and references therein.


\bibitem{footnote1}
Here, a ``non-interacting'' SC means it is described by
its mean-field Hamiltonian even though the SC comes
from some sort of interactions.



\bibitem{Schnyder2008}
A. P. Schnyder, S. Ryu, A. Furusaki, and A. W. W. Ludwig,
Phys. Rev. B \textbf{78}, 195125 (2008).

\bibitem{SRFLnewJphys}
S. Ryu, A. Schnyder, A. Furusaki and A. W. W. Ludwig,
New J. Phys. \textbf{12}, 065010 (2010).


\bibitem{Kitaev2009}
A. Kitaev,
AIP Conf.\ Proc. \textbf{1134}, 22 (2009).

\bibitem{wu} C. Wu, B. A. Bernevig, and S.-C. Zhang, Phys. Rev. Lett. {\bf 96}, 106401 (2006); C. Xu and J. E. Moore, Phys. Rev. B {\bf 73}, 045322 (2006).

\bibitem{Fidkowski-Kitaev}
L. Fidkowski and A. Kitaev, Phys. Rev. B {\bf 81},
134509 (2010).

\bibitem{Fidkowski2}
L. Fidkowski and A. Kitaev, Phys. Rev. B {\bf 83}, 075103 (2011).

\bibitem{Turner}
A. M. Turner, F. Pollmann, and E. Berg, Phys. Rev. B {\bf 83}, 075102 (2011).

\bibitem{Ryu-Moore-Ludwig}
S. Ryu, J. E. Moore, and A. W. W. Ludwig, Phys. Rev. B {\bf 85}, 045104 (2012).

\bibitem{Wang-Qi-Zhang}
Z. Wang, X.-L. Qi, and S.-C. Zhang, Phys. Rev. B {\bf 84}, 014527 (2011).


\bibitem{Chen}
X. Chen, Z.-C. Gu, Z.-X. Liu, and X.-G. Wen,
Phys. Rev. B {\bf 87}, 155114 (2013)..

\bibitem{gu} Z.-C. Gu and X.-G. Wen, arXiv:1201.2648.

\bibitem{XLQ2012}
X.-L. Qi,
New J. Phys. {\bf 15}, 065002 (2013).

\bibitem{RyuZhang2012}
S. Ryu and S.-C. Zhang,
Phys. Rev. B {\bf 85}, 245132 (2012).

\bibitem{footnote4}
Note that
the reflection $\mathcal{R}$
is given by $i\sigma^x$ in the bulk.
When projected on the edge,
$\mathcal{R}$
is given by $i\sigma^y$,
as required by the reality nature of Majorana fermions
on the edge and the condition
$\mathcal{R}^2=-1$.


\bibitem{TeoFuKane2008}
J. C. Y. Teo,
L. Fu,
and
C.\ L.\ Kane,
Phys.\ Rev.\ B \textbf{78}, 045426 (2008).


\bibitem{footnote3}
To explicitly show that the topological classification
in the interacting DIII+R SCs is $\mathbb{Z}_8$,
constructing an adiabatic path connecting the $N=8$ model
to the $N=0$ model without closing the bulk gap should be needed,
as what Fidkowski and Kitaev did in \cite{Fidkowski-Kitaev}
for the (1+1)D BDI SCs,
even though we believe that the classification
argued from the edge consideration is sufficient.


\bibitem{shankar}
R. Shankar,
Phys. Rev. Lett. {\bf 46}, 379 (1981).



\bibitem{Lin-Balents-Fisher}
H.-H. Lin, L. Balents, and M. P. A. Fisher,
Phys. Rev. B {\bf 58}, 1794 (1998).

\bibitem{YaoRyu2012b} C.-K. Chiu, H. Yao, and S. Ryu, arXiv:1303.1843.

\bibitem{Bernard2012}
A systematic study of 1D edge theories of a 2D bulk
was also given recently in
D.\ Bernard,
E.\ -A.\ Kim,
and
A.\ LeClair, Phys. Rev. B
86, 205116 (2012).


\bibitem{Brouwer2005}
See, for example,
P.\ Brouwer, A.\ Furusaki, C.\ Mudry, and S.\ Ryu,
BUTSURI \textbf{60}, 935 (2005) (in Japanese)
(English version: arXiv:cond-mat/0511622).

\bibitem{sato} Y. Ueno, A. Yamakage, Y. Tanaka, and M. Sato, arXiv:1303.0202.

\bibitem{kane} F. Zhang, C. L. Kane, and E. J. Mele, arXiv:1303.4144.

\bibitem{furusaki} T. Morimoto and A. Furusaki, arXiv:1306.2505.

\end{thebibliography}
\end{document}